\documentstyle[aps,preprint
]{revtex}
\begin{document}
% \draft command makes pacs numbers print
\draft
\title{Extraction of bounds on time-reversal non-invariance from neutron
       reactions}
\author{E. D. Davis}
\address{Physics Department, Kuwait University, P. O. Box 5969, Safat, Kuwait}
\author{C. R. Gould}
\address{Physics Department, North Carolina State University, Raleigh, 
         NC 27695-8202\\
         Triangle Universities Nuclear Laboratory, Durham, NC 27708-0308}
% Use byline for 2nd address?
\date{\today}
\maketitle
\begin{abstract}
Ratios involving on-resonance measurements of the three-fold correlation (TC) 
and five-fold correlation (FC) cross-sections for which the dependence on 
some of the unknown spectroscopic data is eliminated are considered. 
Closed-form expressions are derived for the statistical distributions of 
these ratios. Implications for bounds on the variance of matrix elements 
of time-reversal (T) non-invariant nucleon-nucleon (NN) interactions are 
discussed within a Bayesian framework and the competitiveness with bounds
from other experiments is evaluated. The prospects for null FC 
measurements improving by an order of magnitude or 
more upon the current bound on a parity-conserving T-odd interaction are good.
\end{abstract}
% insert suggested PACS numbers in braces on next line

On-resonance three-fold correlation (TC) and five-fold correlation (FC) neutron 
transmission tests of time-reversal invariance~\cite{Ka82,St82,Ba83,Ba86} are 
attractive because of large compound nucleus enhancements in 
sensitivity~\cite{SF80,BG83,BG82,Bu88,GHR90}. Technically, the experiments have 
proved challenging because of the need to eliminate fake signals and the fact 
that they require, in addition to a polarized neutron beam, polarized and 
aligned targets, respectively~\cite{Bo87,Ma94,SMT94,Se94}. However, an aligned 
$^{165}$Ho target for FC 
measurements has been constructed at TUNL~\cite{KGH94} and the first
on-resonance TC measurement with a polarized $^{139}$La target is currently being 
attempted at KEK.

The complexity of compound nucleus (CN) resonances necessitates a statistical
description of on-resonance correlation cross-sections. The first step towards 
a quantitative analysis of on-resonance FC (TC) measurements is the extraction 
of the local root-mean-square (rms) of CN matrix elements of a parity (P)
conserving (non-conserving) T-odd NN
interaction ${\cal V}^{(T)}$ (${\cal V}^{(PT)}$) or, in the event of null 
measurements, a bound on this root-mean-square~\cite{Da89}. To this end, 
it is necessary to introduce a statistical reaction model which relates the 
unknown pure imaginary rms matrix 
element (${\cal V}_{\rm rms}^{(T)}=iv_{\rm T}$ or ${\cal V}_{\rm rms}^{(PT)}=i
v_{\rm PT}$) to the distribution of values of the on-resonance correlation 
cross-sections. In addition, one has to take into account the statistical 
errors in measurements. Since one is dealing an 
intrinsically stochastic observable, the presence of errors can lead to 
non-trivial restrictions on 
the minimum number $M_{\rm min}$ of on-resonance measurements required in 
effect to set a bound on $v_{\rm T}$ and $v_{\rm PT}$~\cite{Da94}.

To date, consideration of the statistics of on-resonance TC and FC 
cross-sections has been limited to the case of measurements at p-wave 
resonances~\cite{BDW90}. In the case of the FC test, measurements at s-wave
resonances are an alternative which can benefit from similar enhancements in
sensitivity~\cite{GHR90} and have the practical advantage that \hbox{s-wave}
resonances of known spin $J$ have been located in $^{166}$Ho whereas no p-wave 
resonances have been found yet. Recently~\cite{HGH98}, it has been suggested 
that it may be advantageous to study at weak s-wave resonances the ratio 
$R_{\rm FD}$ of (in the notation of~\cite{Hn94}) the FC cross-section 
$\sigma_{122}$ to the deformation effect cross-section $\sigma_{022}$ (which 
is of interest in its own right~\cite{DG98}) because dependence on unknown
spectroscopic parameters is partially eliminated. By the same token,
it is appropriate to consider the ratio $R_{\rm TP}$ of the TC cross-section 
$\sigma_{111}$ to the parity-violating cross-section $\sigma_{101}$ at p-wave 
resonances. (For completeness, we note that the cross-sections $\sigma_{kK\Lambda}$ are related to elastic elements $S^J_{l^\prime j^\prime,lj}$ of the scattering matrix by
\[\sigma_{kK\Lambda}=2\pi\lambdabar^2
\frac{\widehat{k}\widehat{K}\widehat{\Lambda}^2}{
\widehat{s}\widehat{I}}{\rm Re}\left[i^{-k-K-\Lambda}
\sum_{Jljl^\prime j^\prime}\widehat{J}^2\widehat{l}\;\widehat{j}\;\widehat{j^\prime}\,
\left<l\Lambda 00|l^\prime 0\right>W(JjIK;Ij^\prime)\left\{
\begin{array}{ccc} l & s & j \\ \Lambda & k & K \\ l^\prime & s & j^\prime  
\end{array}
\right\}\left(\delta_{l,l^\prime}
\delta_{j,j^\prime}-S^J_{l^\prime j^\prime,lj}\right)
\right],\]
where $\lambdabar$ is the reduced wavelength, $\widehat{k}=(2k+1)^{1/2}$ and we adopt standard notations for the Clebsch-Gordan, Racah and 9-j coefficients. Via the optical theorem and a partial wave expansion of the
elastic scattering amplitude, the total cross-section $\sigma_{\rm tot}$ for
neutrons incident on a target nucleus can expressed as a linear combination
of the cross-sections $\sigma_{kK\Lambda}$ given in~\cite{Hn94}.)

In this paper, we discuss the statistics of null measurements of $R_{\rm FD}$ 
and $R_{\rm TP}$. The dominant contribution to the cross-sections 
$\sigma_{122}$ and $\sigma_{111}$ at a resonance arises from the admixture 
under the corresponding symmetry-violating interaction with the {\em nearest\/}
adjacent resonance of the right spin and parity and it has been customary in most previous theoretical studies to
include only this contribution. In fact, this two resonance approximation has 
proved to be very good in the interpretation of parity-violation data. 
Inclusion of the smaller admixtures from other more distant resonances is 
possible either along the lines of the approximate analytical treatment of 
\cite{BDW90} or via Monte Carlo simulations (also conducted in~\cite{BDW90})
employing the random Hamiltonian matrix ensembles underlying statistical 
nuclear theory, but, in both approaches, the results
are rather clumbersome. On the basis of the findings of \cite{BDW90}, it can 
be anticipated that the effect of the admixtures from more distant resonances 
will be to smooth out and broaden the probability density distributions for 
on-resonances values of 
$\sigma_{122}$ and $\sigma_{111}$, enhancing, in particular, the tails of the 
distributions or
the probability of outliers. For the purpose of the present feasibility
study, we choose to ignore such effects, making our estimates conservative.

We show that, in the limit when 
mixing with only the nearest adjacent resonance is taken into account,
simple closed-form expressions for the probability 
densities of $R_{\rm FD}$ and $R_{\rm TP}$ can be derived. Motivated by the 
result of~\cite{Da94}, we then investigate the impact of the size $M$ of a 
sample of measurements of $R_{\rm FD}$ or $R_{\rm TP}$ on the determination
of bounds on rms matrix elements. Our statistical analysis is performed within a Bayesian framework~\cite{SO91,BT73} and constitutes the most
comprehensive yet of on-resonance reaction tests of time-reversal. We end 
by considering the competitiveness of on-resonance TC and FC measurements
with other approaches to constraining the T-odd part of the NN interaction. 
We find that the prospects are good for null FC measurements improving by an 
order of magnitude or more upon the bound on ${\cal V}^{(T)}$ extracted recently from neutron-proton charge symmetry breaking (CSB) experiments~\cite{Si97} (currently, the tightest bound).

At a weak s-wave resonance, $\sigma_{122}$ is dominated by the contribution
arising from the mixing under the interaction ${\cal V}^{(T)}$ of the small 
d-wave component of this resonance (labelled 1 below) with the s-wave component 
of the nearest adjacent s-wave resonance of the same spin $J$ (labelled 2 
below). Retaining only this contribution to $\sigma_{122}$ and approximating 
$\sigma_{022}$ at resonance 1 by its resonant contribution, the ratio 
$R_{\rm FD}\equiv\sigma_{122}\left/\sigma_{022}\right.$ is given by~\cite{HGH98} 
\begin{equation}\label{eq:RFD}
R_{\rm FD}=\sqrt{\frac{2}{3}}\sqrt{\frac{\Gamma_{n2}}{\Gamma_{n1}}}\,z\,
\frac{{\rm Im}\left[{\cal V}^{(T)}_{12}\right]}{E_1-E_2}\equiv{\cal C}_{\rm FD}
\,z\,{\rm Im}\left[{\cal V}^{(T)}_{12}\right],
\end{equation}
where, in terms of the ratio $w\equiv g^{(0)}_{n(3/2)}\left/g^{(0)}_{n(5/2)}
\right.$ of the reduced d-wave neutron partial width amplitudes $g^{(0)}_{n(j)}$ 
of resonance 1 and the ratio of Racah coefficients $\rho(J)\equiv
W(J{\textstyle\frac{1}{2}}I2;I{\textstyle\frac{3}{2}})\left/W(J{\textstyle
\frac{1}{2}}I2;I{\textstyle\frac{5}{2}})\right.$ ($I$ is the target spin),
\begin{equation}\label{eq:zdef}
z\equiv\frac{1+\sqrt{\textstyle\frac{3}{2}}\rho(J)w}{1-\sqrt{\textstyle
\frac{2}{3}}\rho(J)w},
\end{equation}
$E_k$ ($\Gamma_{nk}$) denotes the energy (neutron partial width) 
of resonance $k$ and ${\cal V}^{(T)}_{12}$ denotes the pure imaginary
matrix element of ${\cal V}^{(T)}$ coupling resonances 1 and 2. Using
the known spectroscopic data, we work below with $r_{\rm FD}\equiv 
R_{\rm FD}/{\cal C}_{\rm FD}$,
which, to the extent that the two-level approximation inherent in 
Eq.~(\ref{eq:RFD}) is valid, possesses the ergodicity property required of
observables in statistical nuclear theory~\cite{BFF81}. Unlike $\sigma_{122}$,
$r_{\rm FD}$ (or $R_{\rm FD}$) does not depend on the unobservable d-wave
neutron partial width of resonance 1.

The factor $z$ in Eq.\ (\ref{eq:RFD}) must be characterized statistically
because it involves the ratio of reduced partial width amplitudes $w$ which 
cannot be fixed experimentally. According to the empirically 
grounded Porter-Thomas hypothesis, the reduced amplitudes $g^{(0)}_{n(3/2)}$ 
and $g^{(0)}_{n(5/2)}$ of (spin $J$) s-wave resonances are independent gaussian
random variables with vanishing means and variances given by the running 
averages (over the resonances) $\langle(g^{(0)}_{n(3/2)})^2\rangle$ and 
$\langle(g^{(0)}_{n(5/2)})^2\rangle$, respectively. Thus, the probability 
density function (pdf) of $w$ is the Cauchy distribution~\cite{JK70}
$[\pi\lambda(1+\left[w/\lambda\right]^2)]^{-1}$,
where the scale parameter $\lambda\equiv\sqrt{\langle(g^{(0)}_{n(3/2)})^2
\rangle\left/\langle(g^{(0)}_{n(5/2)})^2\rangle\right.}$. Using the inverse 
relation \hbox{$w=w(z)$} implied by Eq.~(\ref{eq:zdef}), we find that the pdf 
for $z$ 
\begin{equation}\label{eq:pdz}
p_I(z)=\frac{1}{\pi\lambda}\frac{1}{(1+\left[w(z)/\lambda\right]^2)}\left|
\frac{{\rm d}w}{{\rm d}z}\right|
\end{equation}
is also a Cauchy distribution. Below, for simplicity, we use the limiting form
\[p_\infty(z)=\sqrt{\frac{2}{3\pi^2}}\frac{1}{(1+{\textstyle\frac{2}{3}}z^2)}\]
obtained when $I\rightarrow\infty$ and we set $\lambda=1$ (taking advantage of 
the fact that the averages $\langle(g^{(0)}_{n(3/2)})^2\rangle$ and $\langle(
g^{(0)}_{n(5/2)})^2\rangle$ should be approximately equal).

Like the CN matrix elements of a generic two-body interaction~\cite{BFF81},
${\rm Im}\left[{\cal V}^{(T)}_{12}\right]$ in Eq.~(\ref{eq:RFD}) can be assumed 
to be a gaussian random variable of zero mean for different pairs of 
neighbouring s-wave resonances of the same $J$ (and with d-wave components). 
Since the rms value of ${\rm Im}\left[{\cal V}^{(T)}_{12}\right]$ is 
$v_{\rm T}$, $r_{\rm FD}=z\,{\rm Im}\left[{\cal V}^{(T)}_{12}\right]$
has pdf 
\begin{equation}\label{eq:PFD}
P_{\rm FD}(r_{\rm FD})=\frac{1}{v_{\rm T}}f(r_{\rm FD}/v_{\rm T}),
\end{equation} 
where, in terms of the pdf $p(z)$ in Eq.~(\ref{eq:pdz}),
\[f(x)\equiv\frac{1}{\sqrt{2\pi}}\int\limits_{-\infty}^{+\infty}\!p_I(x/y)e^{-
y^2/2}\frac{{\rm d}y}{|y|}.\]
Substituting $p_\infty$ for $p_I$, $f(x)$ evaluates to
\[f_\infty(x)=\frac{1}{\sqrt{3\pi}}\frac{1}{\pi}e^{+x^2/3}E_1(x^2/3),\]
where $E_1$ denotes the exponential integral of order one (defined in 
Eq.~5.1.1 in~\cite{AS64}). A simliar but more complicated expression involving 
$E_1$ holds for $f(x)$ in the general case. We have checked numerically for 
$\lambda=1$ that $f_\infty(x)$ does not differ by more than a percent or so 
from the exact result for $f(x)$.

The dominant contribution to the parity-violating cross-section $\sigma_{101}$
at a p-wave resonance comes from the mixing of this resonance (labelled p below)
via the parity-violating interaction ${\cal V}^{(P)}$ with the nearest s-wave 
resonance of the same spin (labelled s below)~\cite{SF80,BG83}. Likewise,
the dominant contribution to the TC cross-section $\sigma_{111}$ (at the same
p-wave resonance) stems from the mixing of these two resonances under the 
interaction ${\cal V}^{(PT)}$~\cite{BG82}. Keeping only these two 
contributions, the ratio $R_{\rm TP}\equiv\sigma_{111}\left/\sigma_{101}\right.$
reduces to
\[R_{\rm TP}=\sqrt{\frac{2I+1}{2J+1}}\sin(\varphi_{\rm p}+\delta_J)
\frac{{\rm Im}\left[{\cal V}^{(PT)}_{ps}\right]}{x_{1/2}{\cal V}^{(P)}_{ps}}
={\cal C}_{\rm TP}\sin(\phi_{\rm p}+\delta_J){\rm Im}\left[{\cal V}^{(PT)}_{ps}
\right],\]
where, in terms of the neutron partial width amplitudes $g_{n(j)}$ and the
total neutron width $\Gamma_{n{\rm p}}$ of the p-wave resonance, $\varphi_{\rm p}=
\arctan\left(g_{n(3/2)}/g_{n(1/2)}\right)$ and $x_{1/2}=g_{n(1/2)}\!\left/\!\sqrt{
\Gamma_{n{\rm p}}}\right.$, $\delta_J\equiv\arctan\left[\sqrt{2}W(J{\textstyle
\frac{1}{2}}I1;I{\textstyle\frac{3}{2}})\!\left/W(J{\textstyle\frac{1}{2}}I1;I{
\textstyle\frac{3}{2}})\right.\right]$ and ${\cal V}^{(PT)}_{ps}$ (${\cal V}^{
(P)}_{ps}$) denotes the pure imaginary (real) matrix element of ${\cal V}^{
(PT)}$ (${\cal V}^{(P)}$) coupling the s- and p-wave resonances. 

Since the product $x_{1/2}{\cal V}^{(P)}_{ps}$ can be extracted (when the 
two-level approximation holds) from the non-zero value of $\sigma_{101}$ 
measured at the p-wave resonance (but not $x_{1/2}$ and ${\cal V}^{(P)}_{ps}$ 
individually), our candidate for ergodic observable is
$r_{\rm TP}\equiv R_{\rm TP}/{\cal C}_{\rm TP}$.
The angle $\varphi_{\rm p}$ for different p-wave resonances has (under the 
Porter-Thomas hypothesis) a uniform random distribution, implying that $\zeta
\equiv\sin(\varphi_{\rm p}+\delta_J)$ has pdf $(\pi\sqrt{1-\zeta^2})^{-1}$. 
Assuming
Im$\left[{\cal V}^{(PT)}_{ps}\right]$ for different pairs of neighbouring s- and
p-wave resonances (of the same $J$) is a gaussian random variable of zero mean
and rms $v_{\rm PT}$, $r_{\rm TP}=\zeta\,{\rm Im}\left[{\cal V}^{(PT)}_{ps}
\right]$ has pdf 
\[P_{\rm TP}(r_{\rm TP})=\frac{1}{v_{\rm PT}}g(r_{\rm TP}/v_{\rm PT})\] 
in which, using Eq.~3.383.3 of~\cite{GR94},
\[g(x)=\frac{1}{\pi\sqrt{2\pi}}e^{-x^2/4}K_0(x^2/4),\]
where $K_0$ denotes a zero order modified Bessel function (section 9.6 
in~\cite{AS64}). 

We now consider the implications of our results for the extraction of bounds on 
$v_{\rm T}$ and $v_{\rm PT}$ using Bayesian methods. To this end, 
the pdf's $P_{\rm FD}$ and $P_{\rm TP}$ are interpreted as the conditional pdf's 
$P_{\rm FD}(r_{\rm FD}|v_{\rm T})$ and $P_{\rm TP}(r_{\rm TP}|v_{\rm PT})$,
respectively. The likelihood function for a data set $\{r_i\}$ of $M$ values of 
$r_{\rm FD}$ [$r_{\rm TP}$] subject to known experimental errors $\{\sigma_i\}$
is $L\left(\{r_i\}|\{\sigma_i\},v\right)=\prod\limits_{i=1}^Mp(r_i|\sigma_i,
v)$, where $v$ denotes $v_{\rm T}$ [$v_{\rm PT}$] and we assume that the 
(conditional) pdf $p(r_i|\sigma_i,v)$ for the datum $r_i$ is related to 
${\cal P}(r|v)=P_{\rm FD}(r|v)$ [$P_{\rm TP}(r|v)$] by 
\begin{equation}\label{eq:gc}
p(r_i|\sigma_i,v)=\frac{1}{\sqrt{2\pi\sigma_i^2}}\int\limits_{-\infty}^{+
\infty}e^{-(r_i-r)^2/(2\sigma_i^2)}{\cal P}(r|v){\rm d}r.
\end{equation}
(We adopt the notation and terminology of~\cite{SO91}, in which the term 
likelihood and the symbol $L$ are used for both Fisher's likelihood and the
probability of a data sample --- see the footonote on p.~614 of~\cite{SO91}.)
The corresponding Bayesian posterior for $v$ ($=v_{\rm T},v_{\rm PT}$) is of 
the form
\[P(v|\{r_i\})={\cal N}^{-1}L\left(\{r_i\}|\{\sigma_i\},v\right)p(v),\]
where $p(v)$ is the prior distribution for $v$ (which is discussed below) and 
the constant factor $\cal N$ is chosen so that $P(v|\{r_i\})$ is normalized.

For the sake of illustration, we deal with data sets where each $r_i=0$, which 
are conceivably the best for establishing bounds on $v_{\rm T}$ and $v_{\rm 
PT}$. The convolutions in Eq.\ (\ref{eq:gc}) can then  be 
obtained in closed form. Substituting the limiting form $f_\infty$ of $f$ in 
Eq.~(\ref{eq:PFD}), we find (with Eqs.~6.225.1 and 6.225.2 in~\cite{GR94}) that, 
for $v_{\rm T}\not=\sqrt{2/3}\sigma_i$,
\begin{equation}\label{eq:cfFD}
\int\limits_{-\infty}^{+\infty}e^{-r^2/(2\sigma_i^2)}P_{\rm FD}(r|v_{\rm T}){\rm
d}r=\frac{2}{\pi}\frac{\sigma_i}{\sqrt{\left|\sigma_i^2-\frac{3}{2}v_{\rm T}^2
\right|}}{\cal K}(\sqrt{3/2}\,v_{\rm T}/\sigma_i),
\end{equation}
where ${\cal K}(x)\equiv\arccos(x)$~[${\rm arccosh}(x)$] for $x<1$ [$x>1$]; for 
$v_{\rm T}=\sqrt{2/3}\sigma_i$, the integral evaluates to $2/\pi$,
the limiting value of the r.h.s.\ of Eq.~(\ref{eq:cfFD}) in this case. 
The second convolution
\[\int\limits_{-\infty}^{+\infty}e^{-r^2/(2\sigma_i^2)}P_{\rm TP}(r|v_{\rm PT})
{\rm d}r=\frac{2}{\pi}\frac{\sigma_i}{\sqrt{v_{\rm PT}^2+\sigma_i^2}}K\!\left(
v_{\rm PT}^2/[v_{\rm PT}^2+\sigma_i^2]\right)\!,\]
where we have used Eqs.~6.621.3 and 8.113.1 in~\cite{GR94} but taken the 
parametric dependence of the complete elliptic integral of the first kind $K$ 
to be given by Eq.~17.3.1 of~\cite{AS64}. For simplicity, we assume below that 
all errors $\sigma_i$ take on the same value $\sigma_{\rm exp}$. 
 
Making use only of the information that $v_{\rm T}$ and $v_{\rm PT}$ are, by 
definition, non-negative, we take the prior $p(v)\propto\Theta(v)$, where 
$\Theta$ denotes the Heaviside step-function. This choice of improper prior is 
arguably too conservative, but it guarantees that, even for the small data
samples we anticipate, it is the likelihood $L$ which dominates the posterior
$P$ and not the prior $p$ (the ``jury principle'', p.\ 23 of~\cite{BT73}). It 
is because we are able through our choice of prior to rigorously impose the 
constraint that $v_{\rm T}$ and $v_{\rm PT}$ are non-negative that we have 
adopted a Bayesian approach.

In Table \ref{tb:bds}, we summarize the information contained in our posteriors 
$P(v|\{r_i\})$ by quoting the bounds $v^{(\alpha)}$ on $v_{\rm T}$ and $v_{\rm PT}$ 
derived from the relation
\[\int\limits_0^{v^{(\alpha)}}\!P(v|\{r_i\}){\rm d}v=\alpha\]
for three standard choices of the probability $\alpha$ (corresponding to what 
for a gaussian random variable would be bounds at the 1, 2 and 3 $\sigma$ level, 
respectively). Because we work within a Bayesian framework, the bound $v^{(\alpha)}$ 
on $v_{\rm T}$ ($v_{\rm PT}$) has the immediate probabilistic interpretation that 
there is a probability $\alpha$ of $v_{\rm T}$ ($v_{\rm PT}$) lying in the 
interval between 0 and $v^{(\alpha)}$. Complementary information on the mean values 
of $v_{\rm T}$ and $v_{\rm PT}$ 
\[\overline{v}\equiv\int\limits_0^\infty\!vP(v|\{r_i\}){\rm d}v\]
implied by our choice of posteriors is given in Table \ref{tb:mns}. In the limit of 
large $v$, our posterior $P(v|\{r_i\})=O([\ln v/v]^M)$, and so, for $M=1$, it is not 
normalizable and cannot be used to infer bounds $v^{(\alpha)}$, while, for $M=2$, 
there is no finite mean value $\overline{v}$. 

For $M\gg 1$, we find via asymptotic analysis that, to leading order in inverse 
powers of $M$, $\overline{v}_{\rm T}\sim\pi/(\sqrt{6}M)$ and $\overline{v}_{\rm 
PT}\sim 2/\sqrt{\pi M}$; the corresponding asymptotic results for bounds are 
$v^{(\alpha)}_{\rm T}\sim\ln[1/(1-\alpha)]\pi/(\sqrt{6}M)$ and, in terms of the 
inverse of the error function $\rm erf$ (Eq. 7.1.1 in \cite{AS64}), $v^{(\alpha)
}_{\rm PT}\sim 2\,{\rm erf}^{-1}(\alpha)/\sqrt{M}$. The more rapid decrease of 
$\overline{v}_{\rm T}$ and $v^{(\alpha)}_{\rm T}$ with increasing $M$ is already 
appparent for the small values of $M$ in Tables \ref{tb:bds} and \ref{tb:mns}. 
It is plausible that null values of $r_{\rm FD}$ should constrain $v_{\rm T}$ 
more than the same number of null values of $r_{\rm TP}$ would constrain 
$v_{\rm PT}$ (assuming that the experimental errors were identical in both 
instances): $P_{\rm FD}(x)$ falls off less rapidly with increasing $x$ than 
$P_{\rm TP}(x)$.

Our findings on the bounds $v^{(\alpha)}$ are consistent with the general (albeit
less quantitative) assertions of~\cite{Da94}. In the absence of any prior 
information, more than one measurement is necessary to set a bound. With our 
choice of prior one needs at least $M_{\rm min}=4$ measurements to set a 95.5\% 
probability bound comparable in size to the experimental error. It may, perhaps, 
be unwise to attach too 
much significance to this particular value of $M_{\rm min}$ given its dependence 
on our choice of prior and the fact that weaker bounds can, of course, be set 
with fewer measurements. However, it would seem self-defeating if, when one has 
struggled to reduce experimental errors down to some level, one is not able 
to set a high probability bound of the same order of magnitude. Also, for four
(or more) measurements, the likelihood function is sufficiently peaked at small
$v$ ($<\sigma_{\rm exp}$) that, for any prior which is essentially constant 
for $v<\sigma_{\rm exp}$, much the same bounds will emerge. Use of a prior with
this behaviour would seem to be natural for an experiment which seeks to improve 
significantly on the current state of knowledge.

The strength of the interaction ${\cal V}^{(T)}$ is conventionally expressed in
terms of the modulus $\overline{g}_\rho$ of the ratio of the T-odd to T-even 
$\rho$-exchange coupling constants. Bounds on $\overline{g}_\rho$ at the 95\% 
confidence level are $5.8\times 10^{-2}$ from MeV-neutron FC experiments~\cite{HRW97}, 
$9.3\times 10^{-3}$ from 
atomic electric dipole moment (edm) measurements~\cite{HHM94} and $6.7\times 10^{-3}$ 
from neutron-proton CSB experiments~\cite{Si97}. For the 
purpose of order of magnitude estimates~\cite{HHM94,FPS87}, $v_T~\mbox{(in meV)}
\sim 10^3\sqrt{10D}\,\overline{g}_\rho$, where, in the context of the FC measurements 
under discussion, we identify $D$ as the spacing (in eV) between s-wave resonances 
of the same spin. Thus, to improve upon the CSB result, a FC experiment with a 
$^{166}$Ho target ($D=4.6$~eV) will need to establish a 95\% probability 
bound on $v_{\rm T}$ of better than $50$ meV. This would appear to be feasible.
Taking ${\cal C}_{\rm FD}\sim 10\mbox{ eV}^{-1}$ (typical values for weak s-wave 
resonances in $^{166}$Ho range from of order unity to of order 10~\cite{HGH98}), 
four null measurements of $R_{\rm FD}$ with errors of 0.1 will yield a 95\% 
probability bound on $v_{\rm T}$ of about 40 meV; for 12 null measurements 
(with errors of 0.1), the corresponding bound is an order of magnitude smaller. 
Even if the null measurements are somewhat less precise and the values of 
${\cal C}_{\rm FD}$ smaller, it is still possible to match the CSB bound with a 
large enough sample of null measurements. An attractive feature of this experiment 
is that, in the energy interval above neutron threshold for which high neutron 
fluxes are available (up to about 1 keV), there are as many as a couple of 
hundred or so s-wave resonances in $^{166}$Ho which could display a deformation 
effect and hence be used for on-resonance measurements of $R_{\rm FD}$.

By contrast, it will be significantly more difficult for a TC experiment to reduce
the already stringent bound on ${\cal V}^{(PT)}$. Neutron and atomic edm measurements 
imply~\cite{He95} an upper limit of $\sim 10^{-5}$ on the ratio of the coupling 
strengths of P-odd, T-odd $\pi$-exchange (assumed to be the dominant contribution to 
${\cal V}^{(PT)}$) to the coupling strength of P-odd, T-even $\rho$-meson exchange. 
It is reasonable to assume that this upper limit translates into a comparable upper 
limit on the ratio $v_{\rm PT}/v_{\rm P}$ of rms matrix elements ($v_{\rm P}$ denotes 
the rms matrix element of ${\cal V}^{(P)}$). With ${\cal C}_{\rm TP}\sim v_{\rm 
P}^{-1}$ and errors of $1\times 10^{-6}$ in the measurement of $R_{\rm TP}$ (there
exist facilities where a statistical error of this magnitude could conceivably be 
attained), four null measurements of $R_{\rm TP}$ would imply a 95\% probability bound
of $5\times 10^{-6}$ on $v_{\rm PT}/v_{\rm P}$.

% figures follow here
%
% Here is an example of the general form of a figure:
% Fill in the caption in the braces of the \caption{} command. Put the label
% that you will use with \ref{} command in the braces of the \label{} command.
%
% \begin{figure}
% \caption{}
% \label{}
% \end{figure}

% tables follow here
%
% Here is an example of the general form of a table:
% Fill in the caption in the braces of the \caption{} command. Put the label
% that you will use with \ref{} command in the braces of the \label{} command.
% Insert the column specifiers (l, r, c, d, etc.) in the empty braces of the
% \begin{tabular}{} command.
%

\begin{table}
\caption{Bounds $v^{(\alpha)}$ on  $v_{\rm T}$ and $v_{\rm PT}$ in units of the
experimental error $\sigma_{\rm exp}$.}
\label{tb:bds}
\begin{tabular}{*{7}{d}}
            &\multicolumn{3}{c}{$v^{(\alpha)}_{\rm T}$}   
            &\multicolumn{3}{c}{$v^{(\alpha)}_{\rm PT}$}   \\
$M$         &$\alpha=\alpha_1$\tablenotemark[1]&$\alpha=\alpha_2$&$\alpha=\alpha_3$
            &$\alpha=\alpha_1$&$\alpha=\alpha_2$&$\alpha=\alpha_3$\\ \hline
2           &7.12          &$>$100.       &$>$5000.          
            &7.31          &$>$100.       &$>$5000.      \\
3           &1.53          &10.3          &80.4          
            &2.09          &11.4          &81.3          \\
4           &0.786         &3.77          &17.4       
            &1.30          &4.83          &19.8          \\
5           &0.521         &2.12          &7.68       
            &0.995         &3.08          &9.56          \\
6           &0.387         &1.44          &4.52       
            &0.828         &2.31          &6.10          \\
7           &0.308         &1.08          &3.09       
            &0.721         &1.89          &4.48          \\
8           &0.255         &0.860         &2.31       
            &0.645         &1.62          &3.57          \\
9           &0.218         &0.712         &1.83       
            &0.589         &1.43          &2.99          \\
10          &0.190         &0.607         &1.50       
            &0.545         &1.29          &2.59          \\
\end{tabular}
\tablenotetext[1]{$\alpha_1=0.683$, $\alpha_2=0.955$, $\alpha_3=0.997$.}
\end{table}

\begin{table}
\caption{Mean values $\overline{v}$ of $v_{\rm T}$ and $v_{\rm PT}$ in units of 
the experimental error $\sigma_{\rm exp}$.}
\label{tb:mns}
\begin{tabular}{*{9}{d}}
$M$                       &3    &4    &5    &6    &7    &8    &9    &10\\ \hline
$\overline{v}_{\rm T}$    &2.89 &0.997&0.578&0.403&0.308&0.249&0.209&0.180\\
$\overline{v}_{\rm PT}$   &3.33&1.43 &0.968&0.761&0.641&0.563&0.506&0.463\\
\end{tabular}
\end{table}

\end{document}